\documentstyle[12pt,epsfig]{article}
\catcode`\@=11
\long\def\@makefntext#1{
\protect\noindent \hbox to 3.2pt {\hskip-.9pt
$^{{\ninerm\@thefnmark}}$\hfil}#1\hfill}                

\def\@makefnmark{\hbox to 0pt{$^{\@thefnmark}$\hss}}  

\def\ps@myheadings{\let\@mkboth\@gobbletwo
\def\@oddhead{\hbox{}
\rightmark\hfil\ninerm\thepage}
\def\@oddfoot{}\def\@evenhead{\ninerm\thepage\hfil
\leftmark\hbox{}}\def\@evenfoot{}
\def\sectionmark##1{}\def\subsectionmark##1{}}

\renewcommand{\thefootnote}{\fnsymbol{footnote}}

\def\sectionc{\@startsection {section}{1}{\z@}{-3.5ex plus -1ex minus
    -.2ex}{2.3ex plus .2ex}{\bf }}
\def\subsectionc{\@startsection{subsection}{2}{\z@}{-3.25ex plus -1ex minus
   -.2ex}{1.5ex plus .2ex}{\it }}
\renewcommand{\section}[1]{\sectionc{#1}\hspace*{\parindent}}
\renewcommand{\subsection}[1]{\subsectionc{#1}\hspace*{\parindent}}
\newcounter{appendixc}
\newcounter{subappendixc}[appendixc]
\newcounter{subsubappendixc}[subappendixc]
\renewcommand{\thesubappendixc}{\Alph{appendixc}.\arabic{subappendixc}}
\renewcommand{\thesubsubappendixc}
        {\Alph{appendixc}.\arabic{subappendixc}.\arabic{subsubappendixc}}
\renewcommand{\appendix}[1] {\vspace*{0.6cm}
        \refstepcounter{appendixc}
        \setcounter{figure}{0}
        \setcounter{table}{0}
        \setcounter{equation}{0}
        \renewcommand{\thefigure}{\Alph{appendixc}.\arabic{figure}}
        \renewcommand{\thetable}{\Alph{appendixc}.\arabic{table}}
        \renewcommand{\theappendixc}{\Alph{appendixc}}
        \renewcommand{\theequation}{\Alph{appendixc}.\arabic{equation}}
        \noindent{\bf Appendix \theappendixc #1}\par\vspace*{0.4cm}}
\newcommand{\subappendix}[1] {\vspace*{0.6cm}
        \refstepcounter{subappendixc}
        \noindent{\bf Appendix \thesubappendixc. #1}\par\vspace*{0.4cm}}
\newcommand{\subsubappendix}[1] {\vspace*{0.6cm}
        \refstepcounter{subsubappendixc}
        \noindent{\it Appendix \thesubsubappendixc. #1}
        \par\vspace*{0.4cm}}

\def\abstracts#1{{

\centering{\begin{minipage}{13.2truecm}\footnotesize\baselineskip=13pt\noindent
        \parindent=0pt #1
        \end{minipage}}\par}}

\newcommand{\bibit}{\it}
\newcommand{\bibbf}{\bf}
\renewenvironment{thebibliography}[1]
        {\begin{list}{\arabic{enumi}.}
        {\usecounter{enumi}\setlength{\parsep}{0pt}
\setlength{\leftmargin 0.75cm}{\rightmargin 0pt}
         \setlength{\itemsep}{0pt} \settowidth
        {\labelwidth}{#1.}\sloppy}}{\end{list}}

\newcounter{itemlistc}
\newcounter{romanlistc}
\newcounter{alphlistc}
\newcounter{arabiclistc}

\newcommand{\fcaption}[1]{
        \refstepcounter{figure}
        \setbox\@tempboxa = \hbox{\footnotesize Figure~\thefigure. #1}
        \ifdim \wd\@tempboxa > 6in
           {\begin{center}
        \parbox{6in}{\footnotesize\baselineskip=13pt Figure~\thefigure. #1}
            \end{center}}
        \else
             {\begin{center}
             {\footnotesize Figure~\thefigure. #1}
              \end{center}}
        \fi}

\newcommand{\tcaption}[1]{
        \refstepcounter{table}
        \setbox\@tempboxa = \hbox{\footnotesize Table~\thetable. #1}
        \ifdim \wd\@tempboxa > 6in
           {\begin{center}
        \parbox{6in}{\footnotesize\baselineskip=13pt Table~\thetable. #1}
            \end{center}}
        \else
             {\begin{center}
             {\footnotesize Table~\thetable. #1}
              \end{center}}
        \fi}

\def\@citex[#1]#2{\if@filesw\immediate\write\@auxout
        {\string\citation{#2}}\fi
\def\@citea{}\@cite{\@for\@citeb:=#2\do
        {\@citea\def\@citea{,}\@ifundefined
        {b@\@citeb}{{\bf ?}\@warning
        {Citation `\@citeb' on page \thepage \space undefined}}
        {\csname b@\@citeb\endcsname}}}{#1}}

\newif\if@cghi
\def\cite{\@cghitrue\@ifnextchar [{\@tempswatrue
        \@citex}{\@tempswafalse\@citex[]}}
\def\citelow{\@cghifalse\@ifnextchar [{\@tempswatrue
        \@citex}{\@tempswafalse\@citex[]}}
\def\@cite#1#2{{$\null^{#1}$\if@tempswa\typeout
        {IJCGA warning: optional citation argument
        ignored: `#2'} \fi}}
\newcommand{\citeup}{\cite}

\font\twelvebf=cmbx10 scaled\magstep 1
\font\twelverm=cmr10  scaled\magstep 1
\font\twelveit=cmti10 scaled\magstep 1
\font\elevenbfit=cmbxti10 scaled\magstephalf
\font\elevenbf=cmbx10     scaled\magstephalf
\font\elevenrm=cmr10      scaled\magstephalf
\font\elevenit=cmti10     scaled\magstephalf
\font\bfit=cmbxti10
\font\tenbf=cmbx10
\font\tenrm=cmr10
\font\tenit=cmti10
\font\ninebf=cmbx9
\font\ninerm=cmr9
\font\nineit=cmti9
\font\eightbf=cmbx8
\font\eightrm=cmr8
\font\eightit=cmti8

\def\pmb#1{{#1}}
\def\prp#1{\rmb{#1}_\perp}
\def\ie{{\it i.e.,}}
\def\eg{{\it e.g.,}}
\def\etal{{\it et al.}}
\def\vs{{\it vs.}}
\def\etc{{\it etc.}}
\def\half{{\textstyle{1\over2}}}
\def\>{\rangle}
\def\<{\langle}
\def\Poincare{Poincar\'e}
\def\Schrodinger{Schr\"odinger}
\def\rmb#1{{\bf #1}}
\def\fmb#1{{\tilde{\bf #1}}}
\def\mat#1{#1}
\def\dfn{\equiv}
\def\>{\rangle}
\def\E{{\bf E}}
\def\P{{\bf P}}
\def\p{{\bf p}}
\def\J{{\bf J}}
\def\j{{\bf j}}
\def\K{{\bf K}}
\def\k{{\bf k}}
\def\X{{\bf X}}
\def\V{{\bf V}}
\def\v{{\bf v}}
\def\Y{{\bf Y}}
\def\half{{\textstyle{1\over2}}}
\begin{document}

\centerline{\normalsize\bf BARYON CURRENT MATRIX ELEMENTS}
\baselineskip=15pt
\centerline{\normalsize\bf IN A RELATIVISTIC QUARK MODEL}
\vspace*{0.6cm}
\centerline{\footnotesize B. D. Keister}
\baselineskip=13pt
\centerline{\footnotesize\it Department of Physics, Carnegie Mellon University}
\baselineskip=13pt
\centerline{\footnotesize\it Pittsburgh, PA 15213 USA}
\centerline{\footnotesize E-mail: keister@cmu.edu}
\vspace*{0.3cm}
\vspace*{0.3cm}
\centerline{\footnotesize S. Capstick}
\baselineskip=13pt
\centerline{\footnotesize\it Department of Physics, Florida State University}
\baselineskip=13pt
\centerline{\footnotesize\it Tallahassee, FL 32306 USA}
\centerline{\footnotesize E-mail: capstick@scri.fsu.edu}

\vspace*{0.6cm}
\abstracts{Current matrix elements and observables for electro- and
  photo-excitation of baryons from the nucleon are studied in a
  light-front framework. Relativistic effects are examined by
  comparison to a nonrelativistic model and can
  typically be of order 20-25\%, but can be larger for certain
  matrix elements, such as radial transitions
  conventionally used to describe the Roper resonance. A systematic
  study shows that the violation of rotational covariance of the
  baryon transition matrix elements stemming from the use of one-body
  currents is generally small.}
\normalsize\baselineskip=15pt
\setcounter{footnote}{0}
\renewcommand{\thefootnote}{\alph{footnote}}

Much of what we know about excited baryon states has grown out of
simple nonrelativistic quark models of their structure. These models
were originally proposed to explain the systematics in the
photocouplings of these states, which are extracted by partial-wave
analysis of single-pion photoproduction experiments.  Much more can be
learned about these states from exclusive electroproduction
experiments.  Electroproduction experiments measure the $Q^2$
dependence of these form factors, and so simultaneously probe the
spatial structure of the excited states and the initial nucleons. Both
photoproduction and electroproduction experiments can be extended to
examine final states other than $N\pi$, in order to find `missing'
states which are expected in symmetric quark models of baryons but
which do not couple strongly to the $N\pi$ channel.\cite{KI,CR} Such
experiments are currently being carried out at lower energies at
MIT/Bates and Mainz. Many experiments to examine these processes up to
higher energies and $Q^2$ values will take place at TJNAF.

It is clear that, once the momentum transfer becomes greater than the
mass of the constituent quarks, a relativistic treatment of the
electromagnetic excitation is necessary.  However, even at low
momentum transfer, the ratio $p_q/\omega_q$ of the average quark
momentum to its average energy is of order unity, which means that
{\it relativistic effects will be significant in any model which
  describes valence quark degrees of freedom.}

The results reported here\cite{CKa} make use of light-front
Hamiltonian dynamics,\cite{KP} in which the constituents are treated
as particles rather than fields.  It shares with light-front
approaches based upon field theories the property that certain
combinations of boosts and rotations are independent of interactions
which govern the quark dynamics, thus making it possible to perform
relatively simple calculations of matrix elements in which composite
baryons recoil with large momenta.  In addition, we make use of a
complete orthonormal set of basis states, composed of three
constituent quarks, which satisfy rotational covariance.  Such a basis
is the natural starting point for dynamical models using the scheme of
Bakamjian and Thomas.\cite{BT}

A consistent relativistic dynamical treatment of constituent quarks in
baryons involves two main parts.  First, the three-body relativistic
bound-state problem is solved for the wavefunctions of baryons with
the assumption of three interacting constituent quarks. Then these
wavefunctions are used to calculate the matrix elements of one-, two-
and three-body electromagnetic current operators.  The conceptual and
formal background for relativistic, directly interacting quarks is
presented in detail in Ref.~\cite{KP}, and are outlined briefly below. 

For quantum mechanical systems, relativistic invariance is equivalent
to the requirement that there be a consistent set of generators of
unitary transformations of inhomogeneous Lorentz transformations.  For
generators of spatial translations (${\bf P}$), rotations (${\bf J}$),
boosts (${\bf K}$) and time translations ($H$), that requirement is
given by the commutation relations. Those relations common to
Galilean-invariant systems are
\begin{equation}
  [J^j, J^k] = i\epsilon_{jkl} J^l;\quad [P^\mu, P^\nu] = 0
\nonumber
\end{equation}
\begin{equation}
  [J^j, P^k] = i\epsilon_{jkl} P^l;\quad [J^j, K^k] = i\epsilon_{jkl} K^l
\end{equation}
\begin{equation}
  [K^j, P^0] = -i P^j;\quad [J^j, P^0] = 0,
  \nonumber
\end{equation}
while those unique to Lorentz-invariant systems are
\begin{equation}
  [K^j, K^k] = -i\epsilon_{jkl} J^l;\quad[K^j, P^k] = i\delta_{jk} P^0.
  \label{eq:AA}
\end{equation}
An inspection of Eq.~(\ref{eq:AA}) shows that the Hamiltonian is {\it
  necessarily} linked to at least some other generators, e.g., the
boosts ${\bf K}$.  In field theory, the generators are constructed via
the energy-momentum stress tensor using the {\it exact} interacting
fields.  The approach taken here follows that of Bakamjian and
Thomas,\cite{BT} with a direct interaction via a mass operator to
construct consistent set of generators.

In light-front dynamics, the generators are reorganized into seven
non-interacting operators $\{P^+;\prp{P};J^3;K^3\}$,
and three interacting generators $\{P^-;\prp{J}\}$.
The Bakamjian construction consists of a mass operator
\begin{equation}
  M_0 \to M = M_0 + U,
\end{equation}
which determines the interacting generators:
\begin{equation}
  P^- = {M^2 + \prp{P}^2\over P^+};
\end{equation}
\begin{equation}
\prp{J} = {1\over P^+}
\Big\{\rmb{e}_3\times\left[\half(P^+ - P^-)\prp{E} - K^3\prp{P}\right]
+ \prp{P} j^3 + M\prp{j}\Big\}
\end{equation}
For a three-quark system, the non-interacting mass operator is
\begin{equation}
  M_0 = \sum_{i=1}^3 \sqrt{m_i^2 + \rmb{k}_i^2},
\end{equation}
and interactions are added directly to the three-quark system:
\begin{equation}
  M_0 \to M = M_0(\rmb{k}_1, \rmb{k}_2, \rmb{k}_3)
  + U(\rmb{k}_1, \rmb{k}_2, \rmb{k}_3).
\end{equation}
The Capstick-Isgur interaction,\cite{CI} consisting of one-gluon exchange plus
confining terms, satisfies all of the necessary formal requirements
for the interaction $U$.  This choice violates cluster separability,
which is normally a problem for systems of particles which are
individually observable, but not for systems of confined quarks.

The calculations reported here are described in detail in
Ref.~\cite{CKa}.  To calculate the current matrix element between
initial and final baryon states, we expand in sets of free-particle
states:
\begin{eqnarray}
\langle M' j; {\tilde{\bf P}}'\mu' | I^+(0) | M j; {\tilde{\bf P}}\mu\rangle 
&=& (2\pi)^{-18}\int d{\tilde{\bf p}}_1'
\int d{\tilde{\bf p}}_2'
\int d{\tilde{\bf p}}_3'
\int d{\tilde{\bf p}}_1
\int d{\tilde{\bf p}}_2
\int d{\tilde{\bf p}}_3
\sum \nonumber  \\
&&\quad 
\langle M' j'; {\tilde{\bf P}}'\mu' 
| {\tilde{\bf p}}'_1 \mu'_1 {\tilde{\bf p}}'_2 \mu'_2 {\tilde{\bf p}}'_3 \mu'_3\rangle 
\nonumber  \\
&&\quad\times
\langle  {\tilde{\bf p}}'_1 \mu'_1 {\tilde{\bf p}}'_2 \mu'_2 {\tilde{\bf p}}'_3 \mu'_3 |
I^+(0)
| {\tilde{\bf p}}_1 \mu_1 {\tilde{\bf p}}_2 \mu_2 {\tilde{\bf p}}_3 \mu_3\rangle 
\nonumber  \\
&&\quad\times
\langle  {\tilde{\bf p}}_1 \mu_1 {\tilde{\bf p}}_2 \mu_2 {\tilde{\bf p}}_3 \mu_3|
M j; {\tilde{\bf P}}\mu \rangle .
\label{MZAA}
\end{eqnarray}

The electroweak current operator has a cluster expansion similar to
that of its nonrelavistic counterpart:
\begin{equation}
  I^\mu(x) = \sum_{j} I_j^\mu(x) + \sum_{j < k} I_{jk}^\mu(x) + \cdots.
\end{equation}
Two-body currents $I_{jk}$ are required for charge-changing (\eg\
$\pi$ exchange)
and/or non-local interactions, as they are in the nonrelativistic
case, and they are also required for covariance of full current.
In the front form, one- and two-body currents can be grouped
separately.   We compute only the contributions from
one-body matrix elements, and assume that the struck quark carries the
current of a free Dirac particle:
\begin{equation}
\langle {\tilde{\bf p}}'\mu' | I^+(0) |{\tilde{\bf p}}\mu\rangle 
=1\delta_{\mu' \mu}.
\label{qlfme}
\end{equation}

The baryon state vectors are in turn related to wavefunctions as
follows: 
\begin{eqnarray}
\langle  {\tilde{\bf p}}_1 \mu_1 {\tilde{\bf p}}_2 \mu_2 {\tilde{\bf p}}_3 \mu_3|
M j; {\tilde{\bf P}}\mu \rangle 
&=& 
\left|{\partial({\tilde{\bf p}}_1, {\tilde{\bf p}}_2, {\tilde{\bf p}}_3) 
\over \partial({\tilde{\bf P}}, {\bf k}_1, {\bf k}_2)}\right|
^{-{1\over2}}
(2\pi)^3 \delta({\tilde{\bf p}}_1 + {\tilde{\bf p}}_2
+ {\tilde{\bf p}}_3 - {\tilde{\bf P}}) \nonumber \\
&&\quad\times
\langle {\textstyle{1\over2}} {\bar\mu}_1{\textstyle{1\over2}} {\bar\mu}_2 |
s_{12}\mu_{12}\rangle 
\langle s_{12}\mu_{12} {\textstyle{1\over2}}{\bar\mu}_3 | s \mu_s\rangle  \nonumber  \\
&&\quad\times
\langle l_\rho \mu_\rho l_\lambda \mu_\lambda | L \mu_L \rangle 
\langle L \mu_L s \mu_s | j \mu\rangle  \nonumber  \\
&&\quad\times
Y_{l_\rho\mu_\rho}({\hat{\bf k}}_\rho)
Y_{l_\lambda\mu_\lambda}({\hat{\bf K}}_\lambda)
\Phi(k_\rho, K_\lambda) \nonumber  \\
&&\quad\times
D^{({\textstyle{1\over2}}){\dag}}_{{\bar\mu}_1\mu_1}[{\underline R}_{cf}({k}_1)]
D^{({\textstyle{1\over2}}){\dag}}_{{\bar\mu}_2\mu_2}[{\underline R}_{cf}({k}_2)]
\nonumber  \\
&&\quad\times
D^{({\textstyle{1\over2}}){\dag}}_{{\bar\mu}_3\mu_3}[{\underline R}_{cf}({k}_3)],
\label{MZAC}
\end{eqnarray}
The quantum numbers of the state vectors correspond to irreducible
representations of the permutation group.  The spins $(s_{12}, s)$ can
have the values $(0, {\textstyle{1\over2}})$, $(1, {\textstyle{1\over2}})$ and $(1, {3\over2})$,
corresponding to quark-spin wavefunctions with mixed symmetry
($\chi^\rho$ and $\chi^\lambda$) and total symmetry ($\chi^S$),
respectively.\cite{IK}  The momenta
\begin{eqnarray}
{\bf k}_\rho &\equiv& {1\over\sqrt{2}} ({\bf k}_1 - {\bf k}_2); \nonumber  \\
{\bf K}_\lambda &\equiv& {1\over\sqrt{6}} ({\bf k}_1 + {\bf k}_2 - 2{\bf k}_3)
\label{MZAD}
\end{eqnarray}
preserve the appropriate symmetries under various exchanges of
${\bf k}_1$, ${\bf k}_2$ and ${\bf k}_3$.
The set of state vectors formed using Eq.~(\ref{MZAC}) and Gaussian
functions of the momentum variables defined in Eq.~(\ref{MZAD}) is
complete and orthonormal.  Since they are eigenfunctions of the
overall spin, they satisfy the relevant rotational covariance
properties.  Any solution to a relativistic model with three
constituent quarks can be written as a linear combination of these
states.  Thus, current matrix elements in any such model can be
expressed in terms of the basis state coefficients and the matrix
elements between basis state vectors. The use of this orthonormal basis 
allows us to examine the transition form factors for many different 
baryons simultaneously. 

Rotational covariance represents a non-trivial constraint in
light-front dynamics.  It necessitates the existence of two-body
current operators because of the interaction dependence of the
four-vector current.  One can test the extent to which rotational
covariance is violated by constructing a quantity which should vanish
under exact covariance, and comparing it to non-vanishing physical
matrix elements.  Helicity conservation yields the following
constraint: 
\begin{equation}
\sum_{\lambda'\lambda}
D^{j{\dag}}_{\mu'\lambda'}({\underline R}'_{ch})
\langle M'j'; {\tilde{\bf P}}'\lambda' | I^+(0) | M j; {\tilde{\bf P}}\lambda\rangle 
D^j_{\lambda\mu}({\underline R}_{ch}) = 0,\quad |\mu'-\mu| \ge 2,
\label{BAA}
\end{equation}
where
\begin{equation}
{\underline R}_{ch} = {\underline R}_{cf}({\tilde{\bf P}}, M) {\underline R}_y({\pi\over2}), \quad
{\underline R}'_{ch} = {\underline R}_{cf}({\tilde{\bf P}}', M) {\underline R}_y({\pi\over2}),
\label{BAABA}
\end{equation}
and ${\underline R}_{cf}$ is a Melosh rotation.
For elastic scattering from a nucleon, Eq.~(\ref{BAA}) is trivially
satisfied.  For a transition ${1\over2}\to{3\over2}$, there is
a single non-trivial condition, while for ${1\over2}\to{5\over2}$, there are
three.

The result of combining Eqs.~(\ref{MZAA})--(\ref{MZAC}) is a
six-dimensional integral over two relative three momenta. These
integrations are performed numerically, as the angular integrations
cannot be performed analytically.  The integration algorithm is a
multidimensional quadrature technique recently generalized and
extended to higher degree and generalized.\cite{quad}  Uncertainties
are typically on the order of a few percent for the
largest matrix elements.  The light-quark mass is
taken\cite{HISR,CI} to be $m_u=m_d=220$ MeV.

Using the techniques outlined above we can form the light-front
current matrix elements for nucleon elastic scattering $\langle M_N\,
{\textstyle{1\over2}}; {\tilde{\bf P}}'\mu' | I^+(0) | M_N\,
{\textstyle{1\over2}}; {\tilde{\bf P}}\mu\rangle $, from
Eq.(~\ref{MZAA}). We have evaluated Eq.(~\ref{MZAC}) using a simple
ground-state harmonic oscillator basis state, $\Phi_{0,0}(k_\rho,
K_\lambda)= {\rm
exp}\left\{-[k_\rho^2+K_\lambda^2]/2\alpha_{HO}^2]\right\}/(\pi^{\textstyle{3\over2}}\alpha_{HO}^3)$,
where the oscillator size parameter $\alpha_{\rm HO}$ is
taken\cite{IK,KI} to be $0.41$ GeV, and using the (CI) wavefunctions
which result from the full solution of the relativized model mass
operator,\cite{CI} expanded up to the $N=6$ oscillator
shell. Eq.~(\ref{qlfme}) applies equally well to quark spinor and
nucleon spinor current matrix elements, so we can extract $F_1(Q^2)$
and $F_2(Q^2)$ for the nucleons directly from the above light-front
matrix elements.

\begin{figure}
\vspace*{1.5cm}
\vbox{          
\hbox{\hspace*{-2.0cm}
\epsfig{file=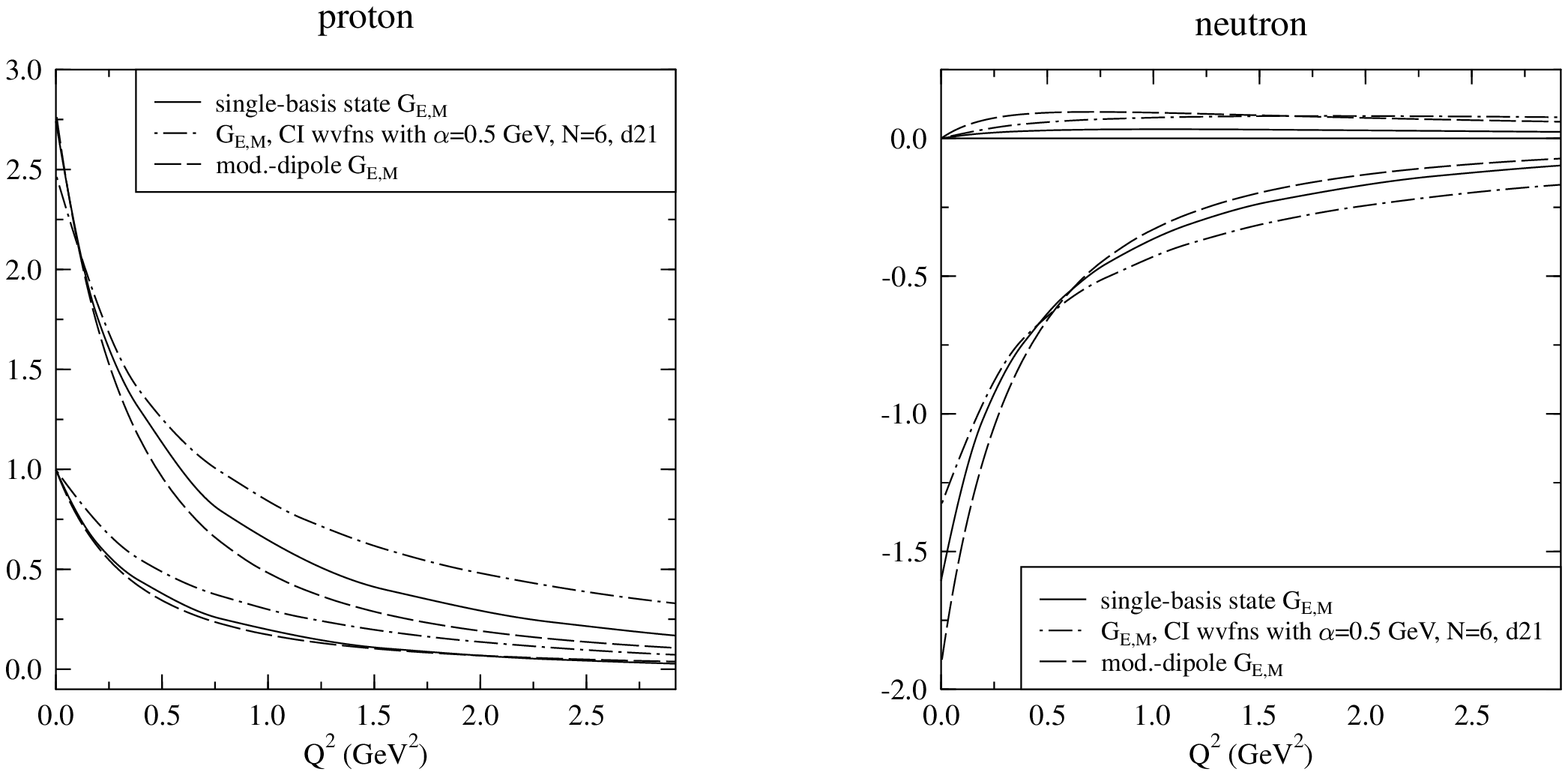,width=18cm,angle=270}}
\vspace*{-8.0cm}
\caption{Proton and neutron elastic form factors $G_E$ and
$G_M$.\label{npGEGM}} } 
\end{figure}

Figure~\ref{npGEGM} compares the proton and neutron $G_E$ and $G_M$
calculated with these two wavefunctions, and the modified-dipole fit
to the data. Our choice of quark mass for the relativistic
calculation, while motivated by previous work,\cite{HISR,CI} gives a
reasonable fit to the nucleon magnetic moments. The single-oscillator
relativistic calculation yields proton charge and magnetic radii close
to those found from the slope near $Q^2$=0 of the dipole fit to the
data. The relativistic calculation using the relativized model
wavefunctions falls off too slowly with $Q^2$, which is due to the
larger probability of higher-momentum components in these
wavefunctions. This confirms the results of previous
work\cite{CPSSnucleon} using these wavefunctions, where the nucleon
form factors were fit by the adoption of relatively soft form factors
for the quarks.

Figure~\ref{pGA} compares the axial-vector form factor $G_A(Q^2)$ and
$G_E(Q^2)$ for the proton calculated with the CI
wavefunctions. Relativistic effects are known\cite{HISR} to reduce
the axial coupling constant $g_A$ from the static nonrelativistic
quark model value of 5/3 to more like the physical value of 1.25 using
simple single-Gaussian wavefunctions; using the CI wavefunctions $g_A$
is reduced further\cite{ChungCoester} due to the higher momenta of
the quarks in these wavefunctions. As expected from the data for
$G_A(Q^2)$, the axial form factor falls with $Q^2$ more slowly than
$G_E$.

Figure~\ref{Delta} shows our relativistic results for the
$A_{\textstyle{1\over2}}$, $A_{\textstyle{3\over2}}$, and
$C_{\textstyle{1\over2}}$ helicity amplitudes for electroexcitation of
the $\Delta{\textstyle{3\over2}}^+(1232)$ from nucleon targets using
$N=6$ CI wavefunctions for the initial and final momentum-space
wavefunctions, compared to relativistic results using the single
oscillator-basis state above. The parameters $\alpha_{HO}$ and
$m_{u,d}$ are the same as above. The relativistic calculation does not
solve the problem of the long-standing discrepancy between the
measured and predicted photocouplings (which are essentially
transition magnetic moments as the transition is almost purely M1),
although the behavior of the single-oscillator relativistic
calculation is similar to the faster-than-dipole fall off found in the
data. Like the nucleon magnetic and axial moments, the photocouplings
are reduced further by the adoption of the CI wavefunctions, and the
form factors drop more slowly with $Q^2$. A reasonable fit to the
$Q^2$ dependence of the data is achieved by Cardarelli {\it et
al.}\cite{CPSSdelta} using the CI wavefunctions and soft quark
form factors which fit the nucleon form factors.

We have also plotted the numerical value of the rotational covariance
condition (multiplied by the normalization factor
$\zeta\sqrt{4\pi\alpha /2K_W}$ for ease of comparison to the physical
amplitudes), given by the left-hand side of Eq.~(\ref{BAA}), for
$\vert \mu^\prime - \mu \vert=2$. At lower values of $Q^2$ the
rotational covariance condition expectation value is a small fraction
of the transverse helicity amplitudes, but approximately the same size
as $C_{\textstyle{1\over2}}$ and larger than the value of $E2/M1$
implied by our $A_{\textstyle{1\over2}}$ and
$A_{\textstyle{3\over2}}$.

Given the controversy surrounding the nature of the baryon states
assigned to radial excitations of the nucleon in the nonrelativistic
model\cite{radexc}, in Figure~\ref{Roper} we compare nonrelativistic
and relativistic calculations, for both proton and neutron targets,
using for the final wavefunction a simple radially excited basis state
which can be used to represent the $P_{11}$ resonance
$N(1440){\textstyle{1\over2}}^+$. For the initial state we have used
the single oscillator-basis ground state wavefunction above.

There are large relativistic effects, with differences between the
relativistic and nonrelativistic calculations of factors of three or
four. Interestingly, the transverse amplitudes also change sign at low
$Q^2$ values approaching the photon point. The large amplitudes at
moderate $Q^2$ predicted by the nonrelativistic model (which are
disfavored by analyses of the available single-pion electroproduction
data\cite{Stoler}) appear to be an artifact of the nonrelativistic
approximation. This disagreement, and that of the nonrelativistic
photocouplings with those extracted from the data for this
state,\cite{Caps} have been taken as evidence that the Roper
resonance may not be a simple radial excitation of the quark degrees
of freedom but may contain excited glue.\cite{ZPL,LBL} The strong
sensitivity to relativistic effects demonstrated here suggests that
this discrepancy for the Roper resonance amplitudes has a number of
possible sources, including relativistic effects.

We also find in the case of proton targets that there is a sizeable
$C^p_{\textstyle{1\over2}}$, reaching a value of about $40-50\times
10^{-3}$ GeV$^{{\textstyle{1\over2}}}$ at $Q^2$ values between 0.25
and 0.50 GeV$^2$, and increasing at lower $Q^2$
values. Correspondingly, there will be a sizeable longitudinal
excitation amplitude.

\begin{figure}
\noindent
\vbox{          
\hbox{\hspace*{-1.0cm}
\epsfig{file=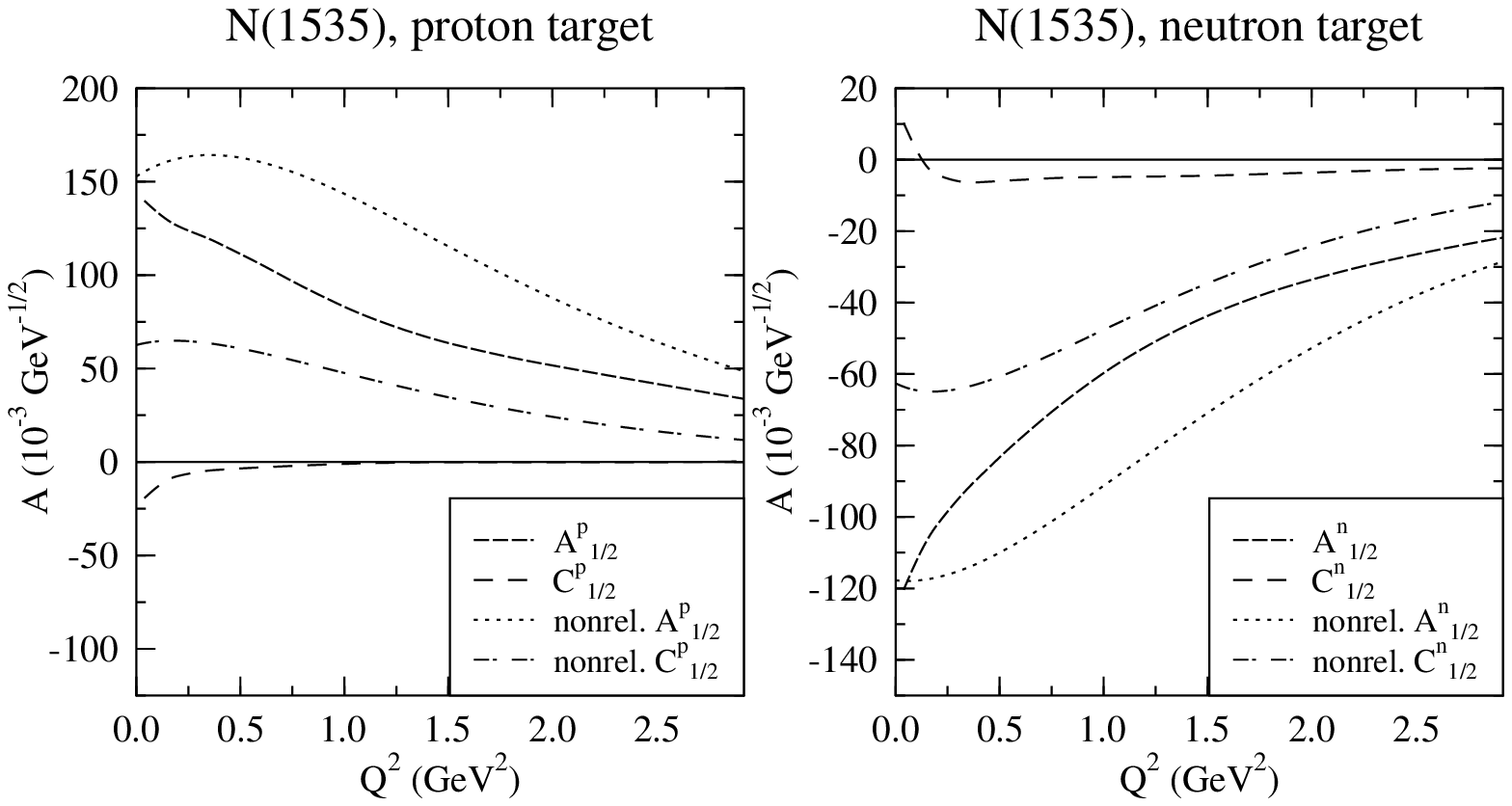,width=16cm}}
\vspace*{-8.0cm}
\caption{Single-basis state relativistic $N(1535)$ electroexcitation
helicity amplitudes, and corresponding nonrelativistic Breit-frame
transverse helicity amplitudes.
\label{N1535}}
}       
\end{figure}

We have also calculated helicity amplitudes for the final state
$N{\textstyle{1\over2}}^-(1535)$, for both proton and neutron
targets. Here we use the same single oscillator-basis state initial
momentum-space wavefunction as above, and final state wavefunctions
which are made up from momentum-space wavefunctions with one or the
other oscillator orbitally-excited. In this case configuration mixing
due to the tensor part of the hyperfine interaction is included in the
final-state wavefunction. Since the two types of orbitally excited
states are degenerate in mass before the application of tensor
spin-spin interactions, they are substantially mixed by them.  The
results for the helicity amplitudes for
$N{\textstyle{1\over2}}^-(1535)$ excitation are compared to the
corresponding nonrelativistic results in Figure~\ref{N1535}.

In contrast to the results shown above, here there is reduced
sensitivity to relativistic effects in the results for the transverse
amplitudes $A_{1/2}$, with the main effect being a hardening of the
$Q^2$ behavior of the transition form factor; this is not the case for
the $C_{1/2}$ amplitudes. For both targets the substantial
nonrelativistic $C_{1/2}$ amplitudes are reduced to essentially zero
in the relativistic calculation.

\section{Discussion and Summary}
The results outlined above establish that there can be considerable
relativistic effects at all values of $Q^2$ in the electroexcitation
amplitudes of baryon resonances, even at $Q^2=0$.  In particular, our
results show that the $Q^2$ dependence of the nonrelativistic
amplitudes is generally modified into one resembling a dipole falloff
behavior, as has been shown in the case of the nucleon form factors.
However, we consider it remarkable that relativistic effects account
for a large part of discrepancy between the nonrelativistic model's
predictions and the physical situation.

Electroexcitation amplitudes of the
$P_{11}$ Roper resonance $N(1440){\textstyle{1\over2}}^+$ and
$N(1710){\textstyle{1\over2}}^+$ states, as well as those of the
$\Delta(1600){\textstyle{3\over2}}^+$, are substantially modified in a
relativistic calculation.  Given the controversial nature of these
states,\cite{ZPL,LBL} we consider this an important result.  Our
results show that relativistic effects tend to reduce the predicted
size of the amplitudes for such states at intermediate and high $Q^2$
values, in keeping with the limited experimental observations for the
best known of these states, $N(1440){\textstyle{1\over2}}^+$.

We have also found that the rotational covariance violation is a small
fraction of the larger amplitudes for the $Q^2$ values considered
here. In cases where the dynamics causes an amplitude to be
intrinsically small, the uncertainty in our results for these
amplitudes becomes larger. In particular, the calculated ratios
$E2/M1$ and $C2/M1$ for the electroexcitation of the
$\Delta(1232){\textstyle{3\over2}}^+$ in the {\it absence} of
configuration mixing of $D$-wave components into the initial and final
state wavefunctions\cite{BDW} are probably 100\% uncertain, and are
thus consistent with zero at all $Q^2$.\cite{e2overm1} This may not
be the case in the presence of such configuration mixing, and we
intend to investigate this possibility, since $\Delta(1232)$
electroproduction is the subject of current experiments at MIT/Bates
and several proposed experiments at CEBAF.\cite{D1232expt}

\vskip12pt

This work was partially supported by the National Science Foundation
under grant PHY-9023586 (B.D.K), by the U.S. Department of Energy
through Contract No. DE-AC05-84ER40150 and Contract No.
DE-FG05-86ER40273, and by the Florida State University Supercomputer
Computations Research Institute which is partially funded by the
Department of Energy through Contract No.  DE-FC05-85ER250000 (S.C.).


\begin{figure}
\vspace*{1.5cm}
\vbox{          
\hbox{\hspace*{1cm}
\epsfig{file=protonga.ps,width=10cm,angle=0}}
\vspace*{-1.5cm}
\caption{Proton axial form factor $G_A$ and $G_E$ calculated with the
CI wavefunctions.\label{pGA}} } 
\end{figure}


\begin{figure}
\noindent
\vbox{          
\hbox{\hspace*{-1.3cm}
\epsfig{file=D1232.ps,width=16cm}}
\vspace*{-6.0cm}
\caption{Single-basis state and CI wavefunction relativistic
$\Delta(1232)$ electroexcitation helicity amplitudes and rotational
covariance condition.\label{Delta}} } 
\end{figure}


\begin{figure}
\noindent
\vbox{          
\hbox{\hspace*{-1.0cm}
\epsfig{file=bcmeN1440.ps,width=16cm}}
\vspace*{-8.0cm}
\caption{Single-basis state relativistic $N(1440)$ electroexcitation
helicity amplitudes, and corresponding nonrelativistic Breit-frame
helicity amplitudes ($C^n_{1/2}$ is zero in the nonrelativistic
model).\label{Roper}} } 
\end{figure}
\vskip 1.0cm


\vspace*{1.0cm}


\begin{thebibliography}{9}
\bibitem{e2overm1} Although it may not
  be obvious from Fig.~\ref{Delta}, our results for the transverse
  amplitudes at $Q^2=0$ yield E2/M1 and C2/M1 which are {\it exactly}
  zero.
\end{thebibliography}
\end{document}